\documentclass[twocolumn,showpacs,pra,superscriptaddress]{revtex4}

\usepackage{amsmath}
\usepackage{amssymb}
\usepackage{graphicx}
\usepackage{epsfig}

\newcommand{\Sc}{Schr\"odinger }
\newcommand{\Heff}{H_{\rm eff}}

\begin{document}

\title{\textbf{Feshbach projection-operator formalism applied to resonance
scattering\\
on Bargmann-type potentials }}

\author{Varvara V. Shamshutdinova$^{1,2}$, Konstantin N. Pichugin$^{2,3}$,
Ingrid Rotter$^2$, and Boris F. Samsonov}

\address{Tomsk State University, 36 Lenin Avenue, 634050 Tomsk,
Russia\\
$^2$Max Planck Institute for the Physics of Complex Systems,
D-01187 Dresden, Germany \\
$^3$Kirensky Institute of Physics, 660036 Krasnoyarsk, Russia }

\begin{abstract}

The projection-operator formalism of Feshbach is applied to
resonance scattering in a single-channel case. The method is based
on the division of the full function space into two segments,
internal (localized) and external (infinitely extended). The
spectroscopic information on the resonances is obtained from the
non-Hermitian effective Hamilton operator $H_{\rm eff}$ appearing
in the internal part due to the coupling to the external part. As
is well known, additional so-called cutoff poles of the $S$ matrix
appear, generally, due to the truncation of the potential. We
study the question of spurious $S$ matrix poles in the framework
of the Feshbach formalism. The numerical analysis is performed for
exactly solvable potentials with a finite number of resonance
states. These potentials represent a generalization of
Bargmann-type potentials to accept resonance states. Our
calculations demonstrate that the poles of the $S$ matrix obtained
by using the Feshbach projection-operator formalism coincide with
both the complex energies of the physical resonances and the
cutoff poles of the $S$ matrix.

\end{abstract}


\maketitle

\section{Introduction}

The Feshbach projection operator (FPO)
formalism~\cite{Feshbach1,Feshbach2} is a powerful method for the
description of resonant scattering and reactions involving light
nuclei \cite{Barz,Rotter_Ann}. In recent years the FPO technique
has been applied to numerous other systems like quantum dots and
microwave cavities \cite{QD1,QD2,QD3,QD4,QD5,QD6,QD7} and atoms in
a laser field \cite{atom1,atom2,atom3}. In its original
formulation, the formalism is based on the introduction of
projection operators~$Q$ and $P$, $QP=PQ=0$,~$P+Q=1$, which
project, respectively, onto the discrete states of a closed system
and the continuous spectrum of a reservoir when their interaction
is neglected. Resonance states then naturally appear as bound
states of the former closed system embedded into a continuum of
open channels, due to the coupling really existing between the
closed system and the reservoir. In other words: the starting
point of the FPO formalism is the assumption that the scattering
event is confined to a certain compact part of the available space
\cite{Fyodorov}. This region constitutes the so-called interaction
region and can be described by the Q subspace. Outside this region
(in the P subspace) the interaction is absent so that the motion
of scattering fragments depends (apart from the total energy $E$)
only on their internal states. Each combination of internal states
of all fragments is called a channel of reaction since it
specifies a set of configurations (depending on $E$) in which the
system can be found long before and long after the scattering
takes places.

The FPO formalism exploits the concept of an effective Hamiltonian
$\Heff$ to describe the open system resulting from the interaction
between the idealized closed system and the reservoir. The
operator $\Heff$ is, naturally, non-Hermitian and depends
explicitly on energy. Its complex eigenvalues $z_\lambda$ are
energy dependent. The solutions of the fixed-point equations for
the eigenvalues provide approximately both energy positions and
inverse lifetimes (widths) of resonance states \cite{Barz}. Using
the FPO formalism, an expression for  the $S$ matrix can be
derived \cite{Rotter_Ann}. It contains the complex eigenvalues
$z_\lambda$ with their full energy dependence. The energy
dependence is important especially in the neighborhood of decay
thresholds and in the regime of overlapping resonances
\cite{Rotter,Okolowicz}.

As it is well known, analytic properties of the $S$ matrix are
very sensitive to both the detailed form of the potential and the
behavior of the potential at infinity. Even at a point where the
potential equals to zero with a computer precision, the truncation
strongly affects the picture, especially the number of the
$S$-matrix poles. In the simple case of the scattering by a
potential of a compact support, i.e., a potential vanishing
outside a given cut off radius, the $S$ matrix has an infinite
number of discrete poles in the lower part of the complex $k$
plane \cite{Humblet,Regge,Newton} whereas an exponential
asymptotic form of the potential can lead to a finite number of
poles (see, e.g., \cite{Faddeev}). This means that for a truncated
potential one obtains not only physical (resonance) $S$-matrix
poles but also so-called cutoff poles~\cite{Meyer}. These poles
are not an artefact; they are correct poles of the truncated
potential but do not cause the characteristic phase shift  by
$\pi$. Therefore there is a need to distinguish between the two
kinds of poles \cite{Meyer}. One way to do so is to use the fact
that the positions of the resonance poles are not affected by
changing the model parameters such as, for instance, the cutoff
radius. From a mathematical viewpoint both types of poles are
correct but they have different origins. This is the reason why
the problem of separation of cutoff poles from the resonance poles
is widely discussed in the literature devoted to the $S$ matrix.

To the best of the authors' knowledge the emergence of spurious
complex eigenvalues has received little attention in the context
of the FPO formalism \cite{Domcke1983,Domcke1986}. We think that
the reason for that may reside in the fact that some authors
(e.g., \cite{Savin}) find the concept of the non-Hermitian
effective Hamiltonian unsuitable in the case of potential
scattering. In our opinion, however, the problem is not
investigated in necessary details and the current paper is just
devoted to fill in this gap.

To avoid unnecessary complications we apply, in the present paper,
the FPO formalism to resonance scattering in its simplest form by
considering only single-channel (elastic) $s$-wave scattering. In
doing so we exclude the appearance of the Feshbach (core excited)
resonance states that are present in the general case (see, e.g.,
\cite{Kukulin}). Thus only  potential (shape) resonances may
appear in our approach. We assume the potential to have a finite
number of resonance states and the continuous spectrum to fill the
positive semiaxis resulting as solutions of the usual radial \Sc
equation. We construct exactly solvable potentials by applying the
method of supersymmetric quantum mechanics (SUSY QM) \cite{Witten}
to the inverse scattering problem (see, e.g., Refs.
\cite{Faddeev,Chadan}). By exact solvability we mean the situation
when solutions of the \Sc equation are available in an explicit
form and, in particular, are expressed in terms of elementary
functions.

It is the authors' opinion that the potentials, the scattering
data of which we set ourselves, represent a good testing ground
for numerous schemes of resonance calculations. Such calculations
show the substantial relevance of the concept of the non-Hermitian
effective Hamiltonian to resonance scattering on a finite range
potential. We show that in this case the FPO formalism gives very
accurate results both for the scattering phase shift and the
positions and widths of physical resonances as well as for the
cutoff poles of the scattering matrix. We discuss the fitness
range of the fixed-point approximation and omit from our
discussions the question of why the estimation of position and
width of the resonance states by this method might be meaningful.
In calculations for concrete reactions by using the FPO method,
this approximation is never used since the $S$ matrix contains the
energy dependent functions~$z_\lambda(E)$. It does {\it not}
contain the energy-independent values that characterize the
positions and widths of the resonance states and are obtained
from, e.g., the solutions of the fixed-point equations
\cite{Rotter,Okolowicz}. In the present paper, we determine the
poles of the $S$ matrix exactly within the FPO formalism and
compare them with the results of the exactly solvable potentials.

The paper is organized as follows. In Sec. \ref{II} we recall the
methods of SUSY quantum mechanics in the context of scattering
theory and construct Bargmann-type potentials supporting resonance
states. Using the standard procedures of quantum scattering theory
in Sec. \ref{III}, we calculate the $S$-matrix poles for truncated
Bargmann-type potentials. In Sec. \ref{IV}, we write down the
basic equations of FPO formalism used in the paper. Our numerical
scheme for the computation is mainly based on the consideration of
quantum scattering on billiards with one attached lead \cite{QD1}.
This model is formulated in the tight-binding approximation
(Anderson model). In Sec. \ref{NR} we present numerical results
obtained by using the two considered methods. In the last section
some conclusions are drawn.

\section{Bargmann-type potentials supporting resonance states
\label{II}}

As it is known, the SUSY approach, when restricted to the
derivation of new exactly solvable quantum problems, is basically
equivalent to the Darboux transformation method (see, e.g., Ref.
\cite{Bagrov}). Therefore we use SUSY and Darboux transformations
as synonyms. The whole class of potentials known as Bargmann-type
potentials (see \cite{Bargmann}) may be obtained from the zero
potential with the help of either usual SUSY transformations or
their confluent forms (see~\cite{SJPA95,AS1,AS2}). Typically, such
a potential has an exponentially decreasing tail and supports a
finite number of bound states and no resonances. Its $S$ matrix
(as well as the Jost function) is a rational function of the
momentum $k$ (see, e.g.,~\cite{Newton}). Any such potential may be
obtained by a proper chain of SUSY transformations with {\it real
factorization constants} (see the next section for details).
Nevertheless, as it is shown in \cite{AIN1}, the use of {\it
complex factorization constants} in higher order transformations
\cite{AIN2} (see also \cite{BS}) may lead to a real potential and
corresponds to an irreducible supersymmetry. The use of such SUSY
transformations permits us to enlarge the class of standard
Bargmann potentials by including potentials supporting resonance
states.

\subsection{Darboux transformation method}

In this section we shortly recall the definition and main
properties of the Darboux transformations method necessary for
subsequent analysis. The interested reader can find a more
detailed exposition
elsewhere~\cite{Bagrov,SJPA95,AS1,AS2,AIN1,AIN2,BS,Junker,Cooper,Bagchi,Bogdan,Samsonov}.

In its pragmatic formulation the method essentially consists in
getting solution $\varphi$ of the (transformed) differential
(Schr\"{o}dinger) equation
\begin{equation}\label{h_1}
    h_1\varphi=E\varphi\,,\quad h_1=-\frac{d^2}{dr^2}+V_1\left(r\right)\,,
\end{equation}
by applying differential transformation operator $L$ of the form
\begin{equation}\label{L}
L=-d/dr+w\left(r\right)\,,
\end{equation}
to a known solution $\psi$ of another (initial) equation
\begin{equation}\label{h_0}
    h_0\psi=E\psi\,,\quad
    h_0=-\frac{d^2}{dr^2}+V_0\left(r\right),
\end{equation}
$\varphi=L\psi$, corresponding to the same value of the parameter
$E$. Here the real-valued function $w\left(r\right)$ called the
superpotential is defined as the logarithmic derivative of a known
solution to Eq. \eqref{h_0} denoted by $u$
\begin{equation}\label{w_u}
    w=u'\left(r\right)/u\left(r\right), \quad h_0u=\alpha u
\end{equation}
with $\alpha\leq E_0$, where $E_0$ is the ground state energy of
the Hamiltonian $h_0$ if it has a discrete spectrum or the lower
bound of the continuous spectrum otherwise. Function $u$ is called
the transformation or factorization function and $\alpha$ is known
as the factorization constant or factorization energy. The
potential $V_1$ is defined in terms of the superpotential $w$ as
\begin{equation}\label{V_1}
    V_1\left(r\right)=V_0\left(r\right)-2w'\left(r\right).
\end{equation}
The knowledge of all two-dimensional solution space
 of the initial equation with a given value of $E\in \mathbb C$
provides the knowledge of all solutions of the transformed
equation corresponding to the same value of $E$. In particular, if
all solutions for $E$ belonging to the spectrum of $h_0$ are known
(so-called physical eigenfunctions of $h_0$) the method provides
us with all physical eigenfunctions of $h_1$.

Since the above procedure does not depend on a particular choice
of the potential $V_0$, the transformed Hamiltonian $h_1$ can play
the role of the initial Hamiltonian for the next transformation
step. In this way one gets a chain of exactly solvable
Hamiltonians $h_0$, $h_1$, \ldots , $h_n$ with the potentials
$V_0$, $V_1$, \ldots , $V_n$. To avoid any confusion we mention
that everywhere, except for especially mentioned cases, we shall
use subscripts to distinguish between quantities related to
different Hamiltonians, $h_0$, $h_1$ , \ldots\ and shall omit them
when discussing general properties regarding all Hamiltonians.

First-order Darboux transformation operator~$L_{j,j+1}$, as
defined by Eq. \eqref{L}, relates solutions of two Hamiltonians
$h_j$ and $h_{j+1}$. If one is not interested in the intermediate
Hamiltonians $h_1$, \ldots , $h_{n-1}$ and all factorization
energies are chosen to be different from each other, the whole
chain may be replaced by a single transformation given by an
$n$th-order transformation operator, denoted by
$L^{\left(n\right)}$, defined as a superposition of $n$
first-order transformation operators. A compact representation of
this operator is given by \cite{Crum}
\begin{eqnarray}\label{L_n}
  \psi_n\left(r,k\right) &=& L^{\left(n\right)}\psi_0\left(r,k\right) \\
   &=& W\left(u_1, \ldots , u_n, \psi_0\left(r,k\right)\right)
   W^{-1}\left(u_1, \ldots , u_n\right)\nonumber,
\end{eqnarray}
where $\psi_0\left(r,k\right)$ is a solution to Eq. \eqref{h_0}
corresponding to the energy $E=k^2$ and $\psi_n\left(r,k\right)$
satisfies
\begin{equation}\label{h_n}
    h_n\psi_n\left(r,k\right)=E\psi_n\left(r,k\right), \quad
    E=k^2.
\end{equation}
The transformation functions $u_j$, although labeled by a
subscript, are eigenfunctions of the initial Hamiltonian
\begin{equation}\label{h_0_u}
    h_0 u_j\left(r\right)=\alpha_j^2u_j\left(r\right).
\end{equation}
These should be chosen in a way that the
Wronskian~$W\left(u_1,\ldots,u_n\right)$ is nodeless and either
real or purely imaginary for $r\in\left(0,\infty\right)$. These
conditions guarantee the absence of singularities in the potential
\begin{equation}\label{V_n}
    V_n=V_0-2\frac{d^2}{dr^2}\ln W\left(u_1,\ldots,u_n\right)
\end{equation}
defining the Hamiltonian $h_n$ of Eq. \eqref{h_n} and its real
character for $r\in(0,\infty)$. In particular, factorization
constants should be either real or come in complex conjugate pairs
with corresponding factorization solutions being either real or in
pairs complex conjugate to each other. Formula \eqref{L_n} is
valid for any $E=k^2$ except for $k=\alpha_j$ $\left(j=1, \ldots,
n\right)$. For these values of~$k$ the corresponding solutions are
\begin{equation}\label{psi_n}
\psi_n\left(r,\alpha_j\right)= W^{\left(j\right)}\left(u_1
,\ldots,u_n\right) W^{-1}\left(u_1 , \ldots,u_n\right)\,,
\end{equation}
where $W^{\left(j\right)}\left(u_1, \ldots , u_n\right)$ is the
$\left(n-1\right)$st order Wronskian constructed from $u_1,\ldots
,u_n$ except for $u_j$, $\left.j=1,\ldots,n\right.$.

\subsection{Jost function for a special chain of
transformations}

Let us choose the following set \cite{Samsonov} of eigenfunctions
of the Hamiltonian $h_0$ \eqref{h_0_u} as transformation functions
for the Darboux transformation of order $2n$:
\begin{equation}\label{u_v}
    u_1\left(r\right), v_1\left(r\right), u_2\left(r\right),
    v_2\left(r\right), \ldots , u_n\left(r\right), v_n\left(r\right),
\end{equation}
\begin{equation}
    h_0u_j\left(r\right)=\alpha_j^2u_j\left(r\right),\quad
    h_0v_j\left(r\right)=\beta_j^2v_j\left(r\right).
\end{equation}
Here  $\alpha$'s and $\beta$'s should be different from each other
and chosen in a way that the corresponding factorization energies,
if they are real, are smaller than the ground state energy of
$h_0$ when it has a discrete spectrum, or less than the lower
bound of the continuous spectrum otherwise. No such restrictions
are imposed on $\alpha$'s and $\beta$'s for complex factorization
energies.

We distinguish between the functions $u$ and $v$ from their
behavior at the origin. The functions $v$ are regular,
$v_j\left(0\right)=0$, and hence are uniquely defined up to a
constant factor, that is not essential for our purpose. The
functions $u_j$ are irregular at the origin,
$u_j\left(0\right)\neq0$, and form the so-called singular family.

In \cite{Samsonov} it was shown that the chain of transformations
with transformation functions \eqref{u_v} transforms the initial
Jost function $F_0\left(k\right)$ of Hamiltonian $h_0$ to the Jost
function,
\begin{equation}\label{F_n}
    F_n\left(k\right)=F_0\left(k\right)
    \prod_{j=1}^n\frac{k-\alpha_j}{k+i b_j}\;, \quad \beta_j\equiv
    ib_j\,,
\end{equation}
corresponding to the  Hamiltonian $h_n$. Since the Jost function
should be analytic in the upper half of the complex $k$ plane (see
Refs.
\cite{Faddeev,Meyer,Domcke1983,Domcke1986,Savin,Kukulin,Witten,Chadan})
all $b$'s must be real and positive. This avoids the appearance of
the so-called redundant poles, which occur as poles of the Jost
function or zeros of the $S$ matrix. Every purely imaginary
$\alpha_j=i a_j$ with $a_j>0$ corresponds to a discrete level
$E_j=-a_j^2$ of $h_n$. No restriction, except the ones discussed
above, is imposed on $\alpha$'s. Complex $\alpha$'s coming in
pairs with real parts of opposite signs may correspond to a
visible resonance. Thus, we say that every pair of complex numbers
$\alpha_j=\pm\mbox{Re}\left[\alpha_j\right]+i\mbox{Im}\left[\alpha_j\right]$
corresponds to a resonance and to mirror resonance states with
complex energy,
\begin{eqnarray}
    E_j^{res}=\left(\pm\mbox{Re}\left[\alpha_j\right]+
    i\mbox{Im}\left[\alpha_j\right]\right)^2=\\
    \left(
    \mbox{Re}\left[\alpha_j\right]^2-
    \mbox{Im}\left[\alpha_j\right]^2\right)\pm
    2i\mbox{Re}\left[\alpha_j\right]\mbox{Im}\left[\alpha_j\right].
\end{eqnarray}
In this case, in accordance with the analytic properties of the
Jost function \eqref{F_n}, the technique developed
in~\cite{Samsonov} remains stable for
$\mbox{Re}\left[\alpha_j\right]<0$, and
$\mbox{Im}\left[\alpha_j\right]<0$. States with
$|\mbox{Re}\left[\alpha_j\right]|>|\mbox{Im}\left[\alpha_j\right]|$
correspond to visible resonances provided
$\mbox{Re}\left[\alpha_j\right]^2-
\mbox{Im}\left[\alpha_j\right]^2$ is small enough~\cite{Roman}.

In the one-channel case the $S$ matrix is a single-valued function
of wave number $k$,
\begin{equation}\label{S_full}
   S=\frac{F\left(-k\right)}{F\left(k\right)}=e^{2i\delta\left(k\right)}.
\end{equation}
In our approach $F_n\left(k\right)$ differs from
$F_0\left(k\right)$ by a rational function of momentum $k$.
Therefore the expression for phase shift $\delta_n\left(k\right)$
becomes rather complicated when the number of transformation
functions is sufficiently large. An alternative expression, which
is more convenient for practical calculations, is \cite{Samsonov}
\begin{equation}\label{delta_n}
    \delta_n=\delta_0-
    \sum_{j=1}^n\arctan\left(\frac{k}{-i\alpha_j}\right)-\sum_{j=1}^n\arctan\left(\frac{k}{b_j}\right).
\end{equation}
In Sect. \ref{NR} we apply the above described technique to obtain
potentials with either one or two resonance states. Moreover,
since our starting potential equals zero, when the usual technique
of Darboux transformations \cite{Samsonov} reproduces Bargmann
potentials \cite{Bargmann}, the potentials from Sec. \ref{NR} are
their generalizations to describe resonance scattering. Below we
will denote them by $V_{Brg}$. It is worthwhile to note that the
solutions of the \Sc equation for these potentials are expressed
in terms of elementary (trigonometric) functions.

\section{$S$-matrix poles for truncated Bargmann potentials
\label{III}}

In order to compare the results obtained by means of the FPO
technique and those obtained by the Darboux method we investigate
the scattering on truncated Bargmann potentials $V_{cut}$, i.e. on
potentials equal to Bargmann potentials $V_{Brg}$ for $r<R_{cut}$,
and equal to zero for $r\geq R_{cut}$. In this case, as it is well
known, the analytic continuation of $S\left(k\right)$ to the
complex $k$ plane is a meromorphic function with infinitely many
poles. In order to calculate their positions and widths we use the
methods of usual quantum scattering theory (see, e.g.,
\cite{Newton,Ballentine}).

For $r\geq R_{cut}$, where $V_{cut}\left(r\right)=0$, the solution
of \Sc equation \eqref{h_1} is a linear combination of plane
waves, which we write as
\begin{equation}\label{Psi_r>R}
    \Psi_{r\geq R_{cut}}=\cos\left(\delta_{cut}\right)\sin\left(kr\right)+
    \sin\left(\delta_{cut}\right)\cos\left(kr\right).
\end{equation}
We denote the phase shift for the truncated potential as
$\delta_{cut}$. The phase shift $\delta_{cut}$ is obtained by
solving Eq. \eqref{h_1} for function $\Psi_{r<R_{cut}}(r)$ in the
region $r<R_{cut}$ and matching it to have form \eqref{Psi_r>R} at
$r=R_{cut}$. Function $\Psi_{r<R_{cut}}(r)$ subject to the
Dirichlet boundary condition $\Psi_{r<R_{cut}}(0)=0$, is uniquely
defined (up to an inessential constant factor). Although at
$r=R_{cut}$ both $\Psi_{r<R_{cut}}$ and $d\Psi_{r<R_{cut}}/dr$
must be continuous, it is sufficient for our purposes to impose
continuity on the logarithmic derivative (see, e.g.,
\cite{Ballentine}),
\begin{eqnarray}\label{}
\gamma &\equiv & \left[\frac{1}{\Psi_{r<R_{cut}}}
\frac{d\Psi_{r<R_{cut}}}{dr}\right]_{r=R_{cut}}
\nonumber \\
&=& \left[\frac{1}{\Psi_{r\geq R_{cut}}}\frac{d\Psi_{r\geq
R_{cut}}}{dr}\right]_{r=R_{cut}}\,,
\end{eqnarray}
which is independent of a multiplicative constant. From here we
find the phase shift of the cutoff potential
\begin{equation}\label{delta_cut}
    \tan\left[\delta_{cut}\left(k\right)\right]=\frac{k\cos\left(kR_{cut}\right)-\gamma\sin\left(kR_{cut}\right)}
    {k\sin\left(kR_{cut}\right)+\gamma\cos\left(kR_{cut}\right)}
\end{equation}
which, upon using Eq. \eqref{S_full}, gives its $S$ matrix
\begin{equation}\label{}
    S_{cut}\left(k\right)=e^{2i\delta_{cut}\left(k\right)}=
    e^{-2ikR_{cut}}\frac{k-i\gamma}{k+i\gamma}\,.
\end{equation}
The poles of $S_{cut}(k)$ are the roots of the transcendental
equation
\begin{equation}\label{transcen}
   \gamma=i k
\end{equation}
which we solve numerically in Sec. \ref{NR}.

\section{Feshbach projection operator approach to potential scattering
\label{IV}}

\subsection{Basic relations of FPO formalism}

As was mentioned in the Introduction, in the FPO formalism
\cite{Feshbach1,Feshbach2} the full function space is divided into
two subspaces: the $Q$ subspace contains all wave functions that
are localized inside the idealized closed system and vanish
outside of it while the wave functions of the $P$ subspace are
extended up to infinity and vanish inside the system; see
\cite{Rotter,Okolowicz}. This division is carried out by using the
projection operators $Q$ and~$P$ ($QP=0=PQ$, $P+Q=1$). The wave
functions of the two subspaces can be obtained by standard
methods: the $Q$ subspace is described by eigenfunctions of
Hermitian Hamiltonian  $H_b$ that characterizes the localized
closed system with a discrete spectrum, while the $P$ subspace is
described by the states of Hermitian Hamiltonian $H_c$, which has
a continuous spectrum. In the FPO formalism, the closed system
becomes open because of a really existing coupling between the
localized closed system and the reservoir, i.e., because of the
coupling between the $Q$ and $P$ subspaces. Due to this coupling,
some discrete states of the closed system become resonance states
of the open system which, in general, have finite life times.

In the present paper we are interested, above all, in the properties
of the effective non-Hermitian
Hamiltonian of the open quantum system, which
acts on the $Q$ subspace and carries the influence of the $P$
subspace. It reads
\begin{equation}\label{H_eff}
    \Heff=H_b+\sum_{c} V_{bc}\frac{1}{E^+-H_c}V_{cb}\,.
\end{equation}
Here, $E^+=E+i\epsilon$ with $\epsilon \to 0$. Further, $V_{bc}$
and~$V_{cb}$ stand for the coupling operators between the $Q$
subspace (described by $H_b$) and the $P$ subspace (environment,
described by $H_c$). The operator $\Heff$ is non-Hermitian and
describes the localized system under the influence of the
reservoir \cite{Okolowicz}.

The non-Hermitian operator $\Heff$ is complex-symmetric and
depends explicitly on energy. Its eigenvalues $z_\lambda$ and
eigenfunctions $\phi_\lambda$,
\begin{equation}\label{z_lambda}
    \left(\Heff-z_\lambda\right)\phi_\lambda=0\,,
\end{equation}
are complex.  The eigenvalues provide not only the energies of the
resonance states but also their widths. The eigenfunctions are
biorthogonal. For more details see \cite{Okolowicz}.

We underline here that values like the $S$ matrix and the cross
section are independent of the manner how the $Q$ and $P$
subspaces are defined. However, in order to obtain  the positions
$E_\lambda$ and widths $\Gamma_\lambda$ of the resonance states
from the eigenvalues  $z_\lambda$ of $H_{\rm eff}$, the two
subspaces have to be defined properly.
Otherwise, the $z_\lambda$ have nothing in common with the
spectroscopic values $E_\lambda - i/2~\Gamma_\lambda$ of the
resonance states. This can be seen, e.g., from the fact that
$\Gamma_\lambda \to 0$ if $Q+P \to Q$.

Using the FPO formalism, the scattering matrix $S$ can  be written
in terms of the effective Hamiltonian and the external scattering
states $\left| E,c\right\rangle$ defined by  $(H_c-E)\left|
E,c\right\rangle=0$. It reads \cite{Rotter_Ann,Okolowicz}
\begin{equation}\label{Smat}
    S_{cc'}=\delta_{cc'}-2\pi i\left\langle E,c\right|V_{cb}
\frac{1}{E-\Heff}V_{bc'}\left| E,c'\right\rangle
    \,.
\end{equation}
Characteristic of Eq. \eqref{Smat} is that it contains $\Heff$
with its explicit energy dependence. The energy dependence of
$\Heff$ plays an important role  near decay thresholds and in the
regime of overlapping resonances. The $S$ matrix \eqref{Smat} is
always unitary.

The FPO formalism may formally be considered as a certain generalization of
the $R$ matrix approach \cite{acta}.  In both cases, the wave
functions of the system are localized in coordinate space ($Q$
subspace in the FPO formalism) and coupled to an extended continuum of
scattering wave functions ($P$ subspace in the FPO formalism).
However, the standard spectroscopic
parameters of the $R$ matrix approach do not contain any
feedback from the continuum of scattering wave functions.
In the FPO formalism, they are replaced by the energy-dependent
functions $E_\lambda$ and $\Gamma_\lambda$ in which the feedback from
the continuum is involved.

We examine the concept of the effective Hamiltonian in connection
with the potential scattering on spherically symmetric potentials.
In order to define the $Q$ subspace that contains the localized
part of the problem, we truncate the potential at a certain cutoff
radius $R_{cut}$. The $P$ subspace is defined then by the
remaining part of the function space being extended up to
infinity. The  operators $V_{bc},~V_{cb}$ describe the coupling
between the two subspaces.  In this paper, we consider a
one-dimensional (1D) quantum  system to which one lead is attached
at a certain point. We will describe such a system in the
framework of the tight-binding approach.

\subsection{One-dimensional tight-binding model for resonance
scattering}

Let us consider the resonance scattering on truncated Bargmann
potentials $V_{cut}$ as described in Secs. \ref{II} and \ref{III}
following the  FPO technique. We choose a radius $R$ such that the
functions defined at~$0\leq r\leq R$ belong to the $Q$ subspace,
while the functions defined at $r>R$ belong to the $P$ subspace.
In order to describe the continuum ($P$ subspace) properly, it
should be $R\geq R_{cut}$.

A common approach to solve the \Sc equation in the context of the
FPO formalism consists of the discretization of the spatial
coordinate. The resulting matrix Hamiltonian is the so-called
tight-binding Hamiltonian (see, e.g., \cite{Datta}) which is
widely used to model electronic transfer in molecules and
condensed matter. To obtain the matrix representation for the
effective Hamiltonian $\Heff$ \eqref{H_eff}, we choose a discrete
lattice whose points are located at~$r=r_i=ia$, $i=1,\ldots,N$
($r_N=R$) and approximate the second order derivative by the
finite differences
\begin{equation}\label{}
    \psi''\approx\frac{1}{a^2}\left\{\psi_{i-1}-2\psi_{i}+\psi_{i+1}\right\}\,,
\end{equation}
where $a=r_{i-1}-r_i$ is a lattice constant (being independent of
$i$). Thus, for $r=r_i$, $i=1,\ldots,N-1$, we obtain the following
finite-difference (or tight-binding) \Sc equation
\begin{equation}\label{recur_rel}
    t\left\{-\psi_{i-1}+2\psi_{i}-\psi_{i+1}\right\}+U_i\psi_i=E\psi_i\,,
\end{equation}
where $U_{i}=V_{cut}(r_i)$
and $t=\frac{1}{a^2}$ is the tight-binding coupling constant.

Effects of scattering are introduced through the coupling between
the box ($Q$ subspace) and a semi-infinite lead ($P$ subspace)
that is attached at the point~$R=r_N$. In order to describe these
effects, we present the solution of the \Sc equation in the lead
as
\begin{equation}\label{Psi_lead}
    \psi_i=e^{-i kr_i}-S_{\rm FPO}\left(k\right)e^{i kr_i}\,,\quad
    i\ge N\,.
\end{equation}
This equation defines the one-channel scattering matrix
$S_{\rm FPO}\left(k\right)$. In the above equation we use the standard
dispersion relation of the tight-binding model,
\begin{equation}
\label{TBdisp} E=t\left[2-2\cos(ka)\right],
\end{equation}
where $a$ is the lattice constant (see above) and $E$ is the real
energy of the system. At $r=r_N$, where the $Q$ subsystem (box) is
coupled to the $P$ subsystem (semi-infinite lead), the \Sc
equation \eqref{recur_rel} takes the form \cite{QD1,ando}
\begin{equation}\label{}
    t\left\{-\psi_{N-1}+(2-e^{i ka})\psi_{N}\right\}=
E\psi_N-2i te^{-i kr_N}\sin(ka)\,.
\end{equation}
Denoting $\Psi=\left(\psi_1, \psi_2, \ldots, \psi_{N-1},
\psi_{N}\right)^T$, we get the  matrix equation
\begin{equation}\label{LSeq}
    \left(E-\Heff \right)\Psi=
\left[E-\left(H_{b}-\tilde{W} \right)\right]\Psi=b\,,
\end{equation}
 with coupling matrix $\tilde{W}_{ij}=\delta_{iN}\delta_{jN}te^{i ka}$ and
\begin{widetext}
\begin{equation}\label{Heff_matrix}
  \Heff=
 \left(%
\begin{array}{cccccccc}
  U_1+2t & -t & 0 & \;\ldots & \;0 & 0 & 0 \\
  -t & U_2+2t & -t & \;\ldots & \;0 & 0 & 0 \\
  0 & -t & U_3+2t & \;\ldots & \;0 & 0 & 0 \\
  \vdots & \vdots & \vdots & \;\ddots & \;\vdots & \vdots & \vdots \\
  0 & 0 & 0 & \ldots & \;-t & \;U_{N-1}+2t & -t \\
  0 & 0 & 0 & \ldots & \;0 & \;-t & -te^{i ka}+2t \\
\end{array}%
\right),\quad b=
\left(%
\begin{array}{c}
  0 \\
  0 \\
  0 \\
  \vdots \\
  0 \\
  2i te^{-ikr_N}\sin(ka) \\
\end{array}%
\right).
\end{equation}
\end{widetext}

Equation \eqref{Heff_matrix} gives us the desired matrix
representation for the effective non-Hermitian Hamiltonian
$\Heff$. The Hamiltonian $\Heff$ obtained from the above pictorial
derivation is completely equivalent to the overall Green function
derived in \cite{Datta}. The matrix equation \eqref{LSeq}
describes the scattering on the discretized 1d quantum system with
Bargmann potential at $r_i< R_{cut}$ and zero potential at $r_i\ge
R_{cut}$ and with $R=R_{cut}=r_N$.

Although the poles of the $S$ matrix are of no relevance for the
scattering and reaction processes in the FPO formalism
\cite{Rotter,Okolowicz}, it is  interesting to estimate their
value. This can be done by using the following standard method
\cite{Barz}. First, the fixed-point equations for the positions of
the resonance states in energy are solved,
\begin{equation} \label{fixed-point_e}
    E_\lambda=\left.\mbox{Re}\left[z_\lambda\right]\right|_{E=E_\lambda},
\end{equation}
and then the widths are defined by
\begin{equation}\label{fixed-point_g}
    \Gamma_\lambda=2\left.\mbox{Im}\left[z_\lambda\right]\right|_{E=E_\lambda}
\; .
\end{equation}
The solutions of these equations provide approximately the
energies  $E_\lambda$ and widths $\Gamma_\lambda$ of the resonance
states as long as the $\Gamma_\lambda$ are small. In the
following, we call the solutions of Egs.\eqref{fixed-point_e} and
\eqref{fixed-point_g} shortened the {\it fixed-point solutions}.

In order to make a meaningful comparison of the results of the FPO
method with those of the exactly solvable potentials, the poles of
the $S$ matrix obtained by using the FPO formalism should be
determined exactly. This can be done  by solving  the  equation
\begin{equation}\label{poles}
    {\rm Det}\left[E-\Heff\left(E\right)\right]=0\,
\end{equation}
in the complex $E$ plane, which follows from the expression
\eqref{Smat}, provided that $V_{bc}$ and $V_{cb}$ do not have
poles. Equation \eqref{poles} differs from an eigenvalue equation
since the effective Hamiltonian depends on energy. We found the
solutions of Eq. \eqref{poles} by determining the intersection of
contour lines for the zero values of the real and imaginary parts
of the determinant. In order to distinguish the solutions of Eq.
\eqref{poles} from the fixed-point solutions,  we call them {\it
$S$ matrix poles} in the following.

Using the $S$ matrix obtained from Eqs. \eqref{LSeq} and
\eqref{Psi_lead} we calculate the phase shift according to its
definition
 \eqref{S_full},
\begin{equation}\label{FPO phase shift}
\delta_{\rm FPO}(k)=-\frac{i}{2}\ln S_{\rm FPO}(k) \, .
\end{equation}

In all our numerical calculations  presented in the next section,
the lattice constant $a$ is chosen to be 0.01. We performed
calculations also with $a=0.001$. These calculations require much
more computer resources and change the results only slightly. The
fixed-point equations are solved for $R=R_{cut}$.

\section{Results}
\label{NR}

\subsection{One-resonance potential}

To illustrate the differences between physical (resonance) poles
and unphysical (cutoff) poles more specifically we first apply the
method described in Sec. \ref{II} to construct a potential with
one resonance state. This allows us to carry out a simultaneous
comparative analysis of the smooth (nontruncated) potential, its
truncated version (both obtained with the help of SUSY technique)
and their cutoff counterpart resulting from the FPO formalism.

Our choice $V_0\left(r\right)=0$ in Eq. \eqref{V_n}, which allows
us to use solutions of the free particle \Sc equation, simplifies
the calculations considerably. A resonance is obtained when two
irregular transformation functions of type \eqref{u_v} with real
parameters $a_1$ and $a_2$ are used in a chain of transformations,
i.e.,
\begin{eqnarray}
   u_1&=&\exp\left(-i\alpha_1r\right)\equiv
   \exp\left(\left(a_1+i a_2\right)r\right),\nonumber\\
   u_2&=&\exp\left(-i\alpha_2r\right)\equiv
   \exp\left(\left(a_1-i a_2\right)r\right)\label{irreg}\,.
\end{eqnarray}
As regular solutions \eqref{u_v} we choose,
 in the current case,
 hyperbolic sine functions
\begin{equation}\label{}
   v_1=\sinh\left(b_1r\right),\quad
v_2=\sinh\left(b_2r\right).\label{reg}
\end{equation}
The real constants $a_i$, $b_i$ should be such that $b_i>0$,
$a_2<a_1<0$. The desired potential follows from Eq. \eqref{V_n}
where we have to calculate a fourth order Wronskian
$W(u_1,u_2,v_1,v_2)$. The calculations are simplified if we notice
that this fourth order Darboux transformation is equivalent to a
set of two first order transformations and one second order
transformation. If we choose for the first transformation the
transformation function $u=u_1$, it results, according to Eqs.
(\ref{w_u}) and (\ref{V_1}), in the zero potential difference
(i.e., the initial potential remains  the potential of the free
particle). Then we have to change only the form of the three other
transformation functions. These functions become $L_1u_2$,
$L_1v_1$, $L_1v_2$ with $L=L_1$ given by Eq. (\ref{L}) where
$w=-i\alpha_1$. Evidently, the function $L_1u_2$ is, up to an
inessential constant factor,  the same exponential as $u_2$, and
the functions $L_1v_1$, $L_1v_2$ become proportional to hyperbolic
cosines with shifted arguments. For the second transformation we
choose the transformation function $L_1u_2$ which still does not
add anything to the zero potential but changes the transformation
functions $L_1v_1\to L_2L_1v_1$ and $L_1v_2\to L_2L_1v_2$ [here
$L_2$ is given by the same formula (\ref{L}) with $w=-i\alpha_2$],
producing additional shifts in their arguments. After that we
realize the second order transformation with the
 transformation functions
\begin{equation}
L_2L_1v_1\sim\sinh\left(b_1r-\zeta_1\right),\quad
L_2L_1v_2\sim\sinh\left(b_2r-\zeta_2\right),
\end{equation}
where
\begin{equation}\label{zetai}
 \tanh  \zeta_i=\frac{2a_1b_i}{b_i^2+a_1^2+a_2^2}\;,\quad
   i=1,2
\end{equation}
which gives the desired one-resonance potential
\begin{widetext}
\begin{equation}\label{V_one res}
   V_{1 res}=\frac{2\left(b_1^2-b_2^2\right)\left[b_2^2\sinh\left(b_1r-\zeta_1\right)^2-
   b_1^2\sinh\left(b_2r-\zeta_2\right)^2\right]}
   {\left[b_2\sinh\left(b_1r-\zeta_1\right)\cosh\left(b_2r-\zeta_2\right)-
   b_1\sinh\left(b_2r-\zeta_2\right)\cosh\left(b_1r-\zeta_1\right)\right]^2}\;.
\end{equation}
\end{widetext}

\begin{figure}[t]
\begin{center}
\begin{minipage}{7.5cm}
\epsfig{file=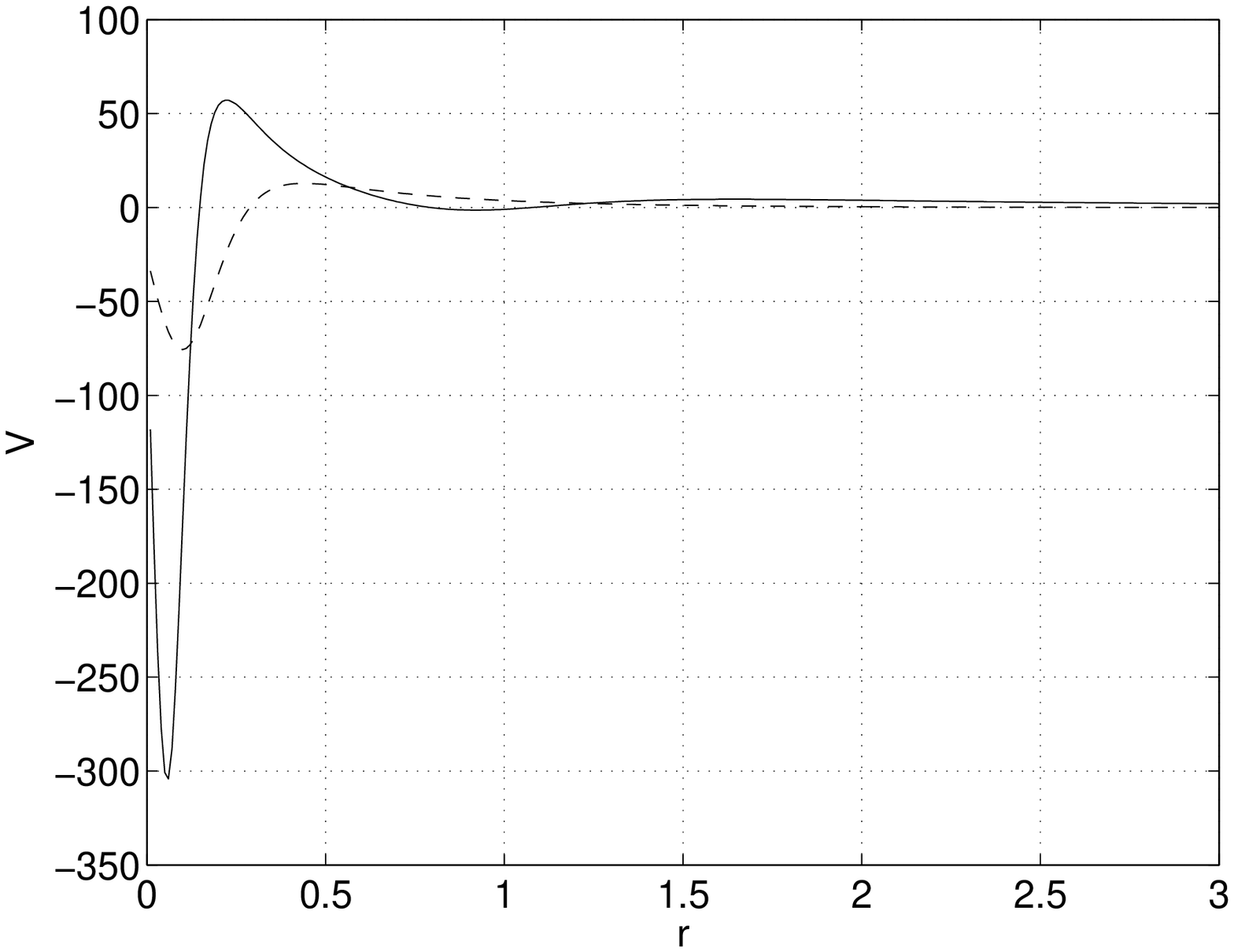, width=7cm} \caption{One-resonance
Bargmann-type potential (dashed line) at $a_1=-0.1$, $a_2=-2$,
$b_1=1$, and $b_2=2$. Two-resonances potential (solid line) at
$a_1=-0.1$, $a_2=-2$, $a_3=-0.08$, $a_4=-3$, $b_1=0.2$, $b_2=0.1$,
$b_3=0.08$, and $b_4=0.05$.} \label{fig1}
\end{minipage}
%
\begin{minipage}{7.5cm}
\epsfig{file=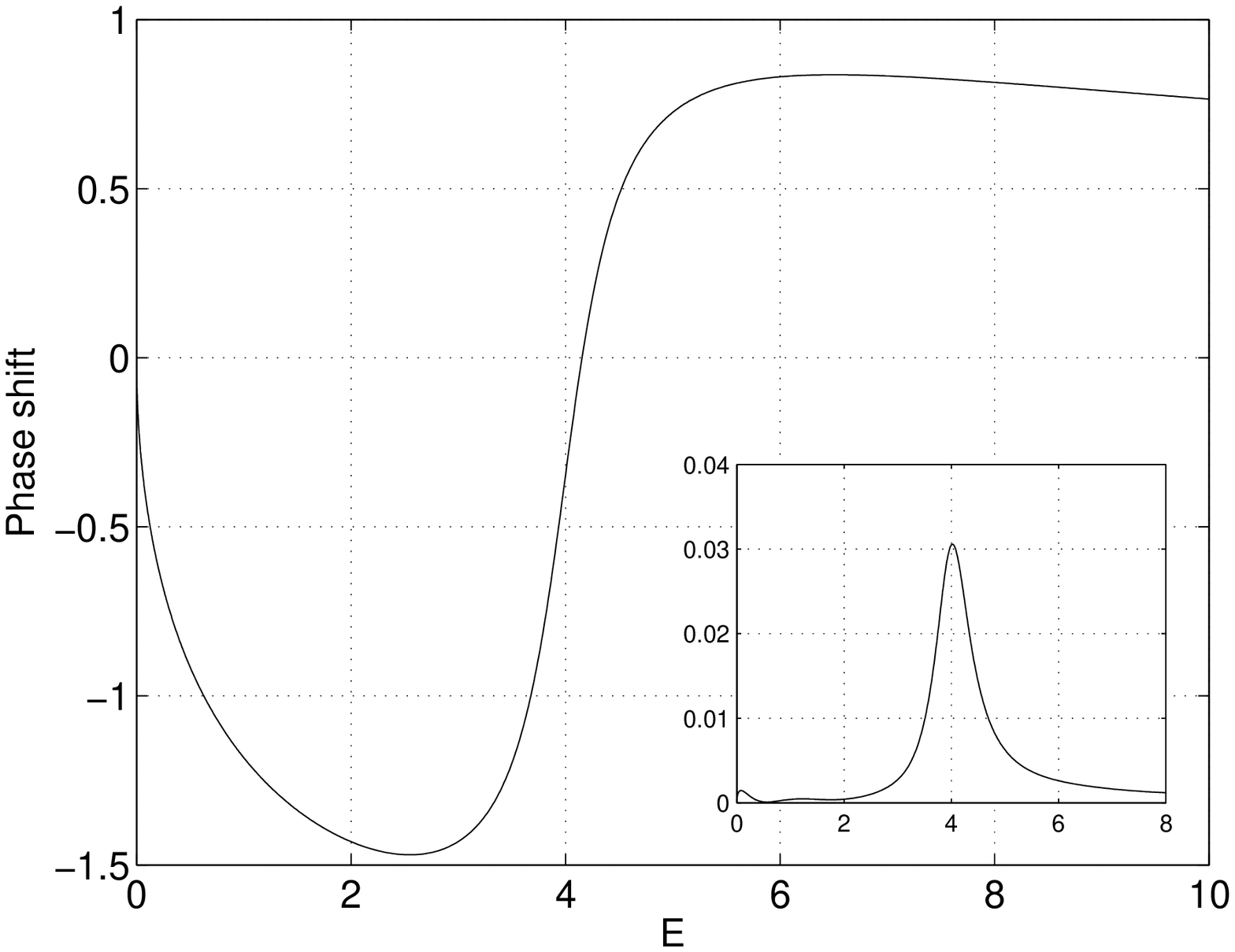, width=7cm} \caption{Phase shift for
one-resonance potential at $a_1=-0.1$, $a_2=-2$, $b_1=1$, $b_2=2$,
and $R_{cut}=5$. Inset shows the difference between
$\delta_{1res}$ and $\delta_{\rm FPO}$. } \label{fig2}
\end{minipage}
\end{center}
\end{figure}

The potential $V_{1res}$ (see Fig. \ref{fig1}, dashed line)
represents a generalization of a two-soliton potential defined on
the positive semiaxis \cite{Matveev}. Instead of two discrete
levels, present in the two-soliton potential, the potential
\eqref{V_one res} has one resonance state.

The Jost function \eqref{F_n} reads
\begin{equation}\label{}
    F_{1res}=\frac{(k-\alpha_1)(k-\alpha_2)}{(k+i b_1)(k+i b_2)}\;.
\end{equation}
Hence the resonance occurs at $k=\alpha_1=-a_2+i a_1$ with the
mirror pole at $k=\alpha_2=a_2+i a_1$, $a_i<0$, $i=1,2$.

The phase shift \eqref{delta_n} is now given by
\begin{equation}\label{delta_V_1}
    \delta_{1res}=-\arctan\frac{2a_1k}{a_1^2+a_2^2-k^2}-
    \arctan\frac{k\left(b_1+b_2\right)}{b_1b_2-k^2}\;.
\end{equation}

To be able to compare the results obtained by the two different
methods, we draw all figures in the $E$  plane [$E=k^2$ for
continuum and Eq. \eqref{TBdisp} in the tight-binding case].

First we compare the scattering phase shift \eqref{FPO phase
shift} calculated by means of FPO technique with that calculated
by Eq. \eqref{delta_V_1}; Fig. \ref{fig2}. The agreement of Eq.
\eqref{FPO phase shift} with the exact result \eqref{delta_V_1}
holds true up to high energy values and in a large range of cutoff
radii. Even in the middle of tight-binding conductance band
$E=2/t$, where we cannot expect good agreement between continuous
and tight-binding models, the difference
$\delta_{1res}-\delta_{\rm FPO}$ does not exceed $0.1$.  The inset
of Fig. \ref{fig2} shows the difference between the values of the
exact phase shift $\delta$ and the numerical $\delta_{\rm FPO}$.
The maximum difference occurs near the resonance position.

According to Eq. (\ref{delta_V_1}), the phase shift is a sum of
two terms. When the parameter values are those used in Fig.
\ref{fig2}, both terms contribute with comparable weight even in
the very neighborhood of the resonance. For other parameter values
(e.g., $a_1=-0.01$, $a_2=-2$ for the resonance term and $b_1=100,
~b_2=200$ for the background), one term dominates in the
neighborhood of the resonance and the phase shift is the standard
one (i.e., $\pi$) in this energy region.

Let us now analyze the calculated spectroscopic data. As shown in
\cite{Nussenzweig}, a cutoff potential produces a  chain of poles
of the $S$ matrix and there are no poles lying below this  chain
in the complex plane.  A narrow physical resonance of the
nontruncated potential separates from the  chain of cutoff poles.
It is chosen to lie close to the real energy axis.
\begin{figure}[h]
\begin{center}
\begin{minipage}{7.5cm}
\epsfig{file=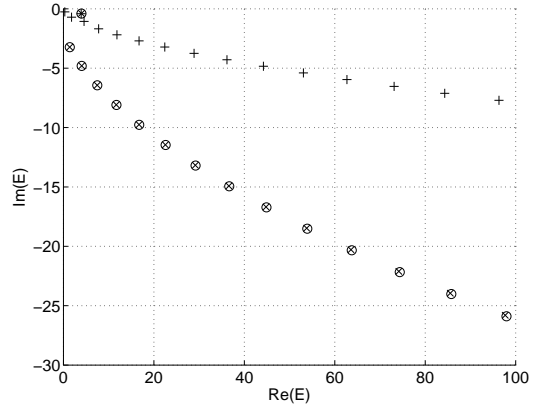, width=7.cm} \caption{Poles
of the truncated one-resonance potential at $a_1=-0.1$, $a_2=-2$,
$b_1=1$, $b_2=2$ and cutoff radius $R_{cut}=5$. The symbols
correspond to the roots of transcendental equation
\eqref{transcen} $(\circ)$, the poles of the $S$ matrix obtained
from Eq. \eqref{poles} $(\times)$, and the fixed-point
approximation \eqref{fixed-point_e}, \eqref{fixed-point_g} $(+)$.
} \label{fig3}
\end{minipage}
\end{center}
\end{figure}

\begin{figure}[]
\begin{center}
\begin{minipage}{7.5cm}
\epsfig{file=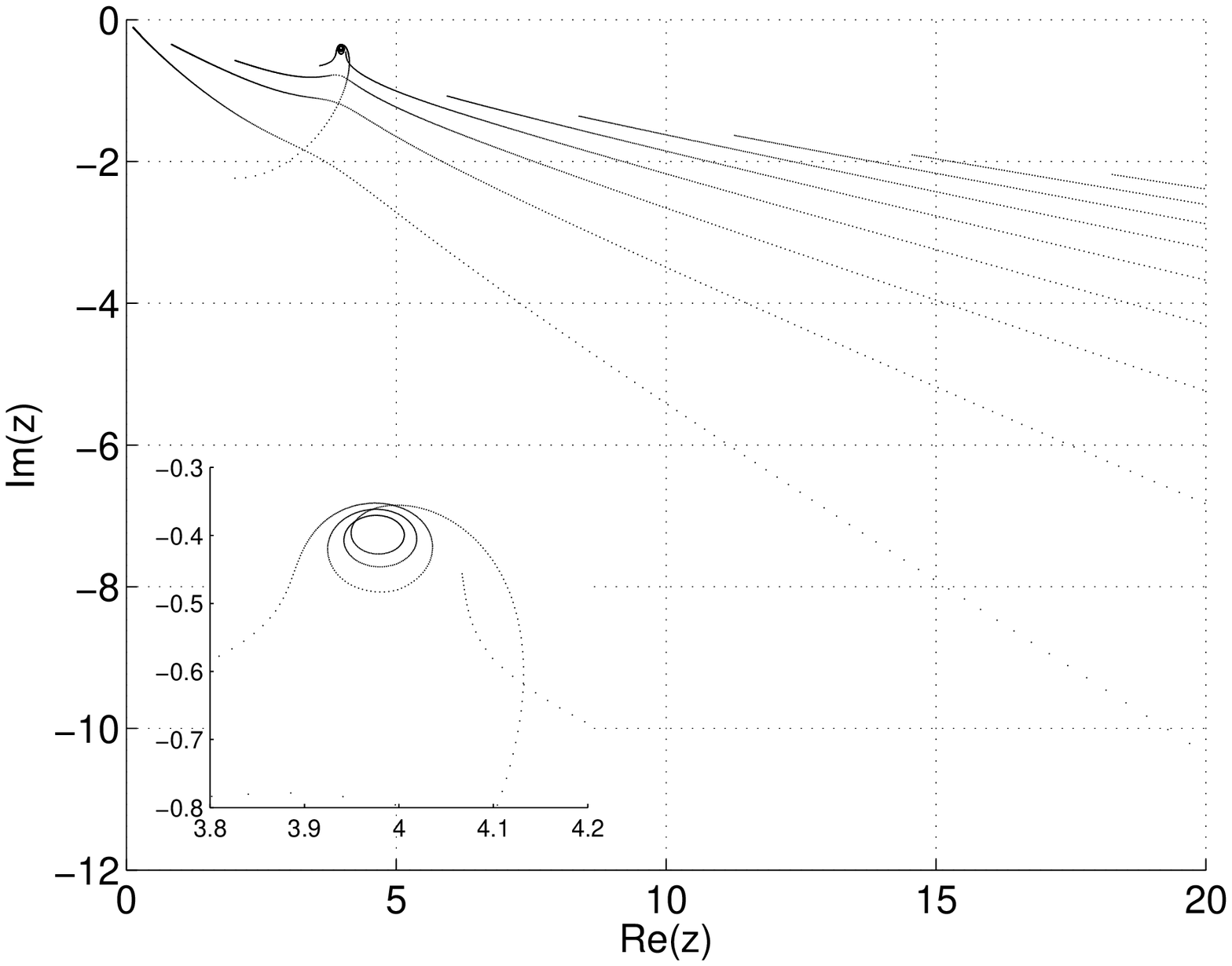, width=7cm}
\caption{Solutions of
fixed-point equations \eqref{fixed-point_e} and
\eqref{fixed-point_g}
 for one-resonance potential at $a_1=-0.1$,
$a_2=-2$, $b_1=1$, and $b_2=2$. Cutoff radius $R_{cut}$ changes
from 0.5 to 7 with step 0.01. With increasing $R_{cut}$, the
trajectories move to small $\mbox{Re}(z)$,~$|\mbox{Im}(z)|$ (with
the exception of the spiralling trajectory).} \label{fig4}
\end{minipage}
%
\begin{minipage}{7.5cm}
\epsfig{file=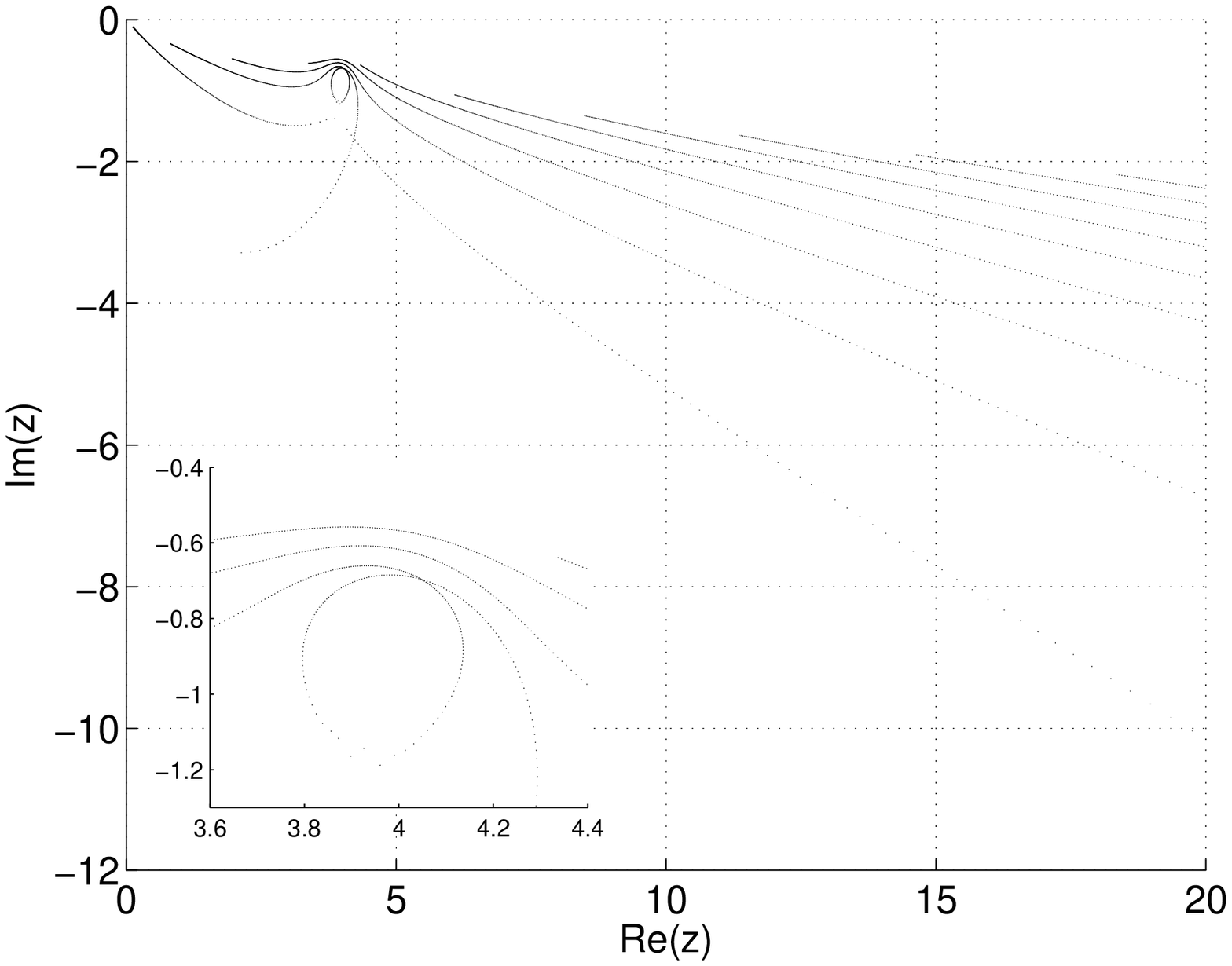, width=7cm} \caption{Solutions of
fixed-point equations \eqref{fixed-point_e} and
\eqref{fixed-point_g} for one-resonance potential at $a_1=-0.2$,
$a_2=-2$, $b_1=1$, and $b_2=2$. Cutoff radius $R_{cut}$ changes
from 0.5 to 7 with step 0.01. With increasing $R_{cut}$,  the
trajectories move to small $\mbox{Re}(z)$,~$|\mbox{Im}(z)|$ (with
the exception of the spiralling trajectory).} \label{fig5}
\end{minipage}
\end{center}
\end{figure}

Figure \ref{fig3} shows all poles ($\mbox{Re}[E]<100$) generated
by the potential \eqref{V_one res} truncated at $R_{cut}=5$. The
roots of transcendental equation \eqref{transcen} are shown as
circles, the crosses present the distribution of complex solutions
of Eq. \eqref{poles}, and daggers stand for the results of the
fixed-point approximation \eqref{fixed-point_e},
\eqref{fixed-point_g}. The resonance which is due to the
nontruncated potential \eqref{V_one res} is clearly separated from
the cutoff poles. Its exact value is $E=\left(-a_2+i
a_1\right)^2=3.99-i0.4$. The fixed-point approximation
\eqref{fixed-point_g} gives the width of the poles with a
significant error. The $S$ matrix pole \eqref{poles} reproduces
this value to a high precision (relative error is less than~$1\%$
for shown poles).

The picture related to the cutoff poles is, according to Fig.
\ref{fig3}, the following: the $S$ matrix poles \eqref{poles}
coincide with the roots of transcendental equation
\eqref{transcen}, whereas the widths \eqref{fixed-point_g}
obtained by using the fixed-point approximation are far too small.
That means that all poles of the $S$ matrix are very good
reproduced by the solutions of Eq. \eqref{poles}. The fixed-point
approximation \eqref{fixed-point_e} allows one to accurately
determine only the position of narrow resonances. This result
agrees with the definitions \eqref{fixed-point_e} and
\eqref{fixed-point_g} according to which the fixed-point equation
is solved only for the real energy.

In order to see how the resonance of the nontruncated potential
separates from the chain of cutoff poles one can consider the
dependence of the pole location on the cutoff radius $R_{cut}$.
Indeed, only the physical resonance is almost independent of
$R_{cut}$. All the other poles move if the cutoff radius is
changed.

To show this dependence in detail we present the results from a
set of calculations by using the FPO technique  for different
cutoff radii $R_{cut}$. First we consider the fixed-point
approximation (Figs. \ref{fig4} and \ref{fig5}). The trajectory of
one of the eigenvalues has a strongly pronounced bight. The
eigenvalue trajectory is spiralling around the correct value of
the physical resonance ($E=3.99-i0.4$), see Fig. \ref{fig4}. For
the broader resonance ($E=3.96-i0.8$) shown in Fig. \ref{fig5} the
spiralling trajectory has less rotations and the localization of
the resonance value becomes more difficult. As a result, the
fixed-point approximation  correctly indicates the position of the
narrow resonance  only.

\begin{figure}[]
\begin{center}
\begin{minipage}{7.5cm}
\epsfig{file=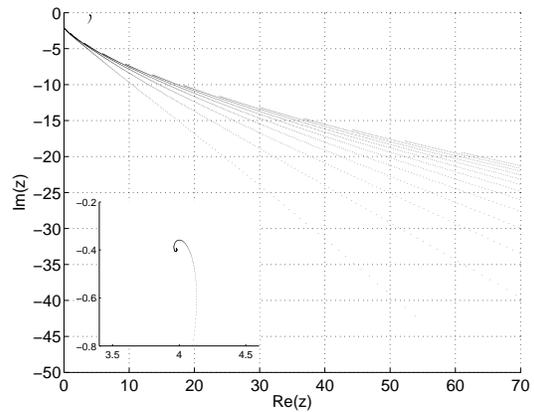, width=7.cm}
\caption{Trajectories of $S$-matrix poles \eqref{poles} for
one-resonance potential at $a_1=-0.1$, $a_2=-2$, $b_1=1$, and
$b_2=2$. Cutoff radius $R_{cut}$ changes from 0.5 to 7 with step
0.01. With increasing $R_{cut}$, the trajectories move to small
$\mbox{Re}(z)$, $|\mbox{Im}(z)|$ (with the exception of the
spiralling trajectory).} \label{fig6}
\end{minipage}
\end{center}
\end{figure}

Figure \ref{fig6} shows the solutions of Eq. \eqref{poles}. They
show a similar dependence on $R_{cut}$ as the fixed-point
solutions (Figs. \ref{fig4} and \ref{fig5}). However, the $S$
matrix poles obtained by solving both Eqs. \eqref{poles} and
\eqref{transcen} coincide. The trajectory of one of the poles
converges very quickly to the resonance value along the spiralling
trajectory.

The results shown in Figs. \ref{fig4}-\ref{fig6} demonstrate that
the cutoff trajectories of the fixed-point solutions
\eqref{fixed-point_e}, \eqref{fixed-point_g} and of the $S$ matrix
poles \eqref{poles} depend strongly on $R_{cut}$. The spiralling
trajectories of the physical resonances arise from a lot of
avoided crossings of the trajectories of neighboring cutoff
trajectories. On this account, the spectroscopic values of the
physical resonances are influenced only a little by varying
$R_{cut}$. The resonance location in Fig. \ref{fig6} is stable
with $R_{cut} \to 7$ in contrast to those in Figs. \ref{fig4} and
\ref{fig5}.

\subsection{Two-resonance potential}

Let us now consider the more complicated case of a potential
supporting two resonance states. To construct the Bargmann-type
potential (see Sec. \ref{II}) we are using eight transformation
functions. As four irregular solutions from the set \eqref{u_v} we
choose two functions \eqref{irreg} with $a_2<a_1<0$ and the
following two functions:
\begin{eqnarray}
      u_3&=&\exp\left(-i\alpha_3r\right)\equiv
   \exp\left[\left(a_3+i a_4\right)r\right],\nonumber\\
   u_4&=&\exp\left(-i\alpha_4r\right)\equiv
   \exp\left[\left(a_3-i a_4\right)r\right]
\end{eqnarray}
 with  $a_3<a_4<0$.
As four regular solutions we choose two
 functions \eqref{reg} and
\begin{equation}\label{}
   v_3=\sinh\left(b_3r\right),\quad
v_4=\sinh\left(b_4r\right)
\end{equation}
with all $b_i>0$. The desired potential follows from Eq.
\eqref{V_n} where this time an eighth order Wronskian should be
calculated. To simplify calculations we replace, similar to the
previous section, the eighth order Darboux transformation by a
chain of four first order transformations, involving exponential
transformation functions only, and one fourth order
transformation. The first order transformations keep unchanged the
zero initial potential but affect the hyperbolic transformation
functions, producing only shifts in their arguments. Thus the
potential is calculated by Eq. \eqref{V_n} with $V_0=0$ and the
fourth order Wronskian $W(\tilde v_1,\ldots,\tilde v_4)$. Here
$\tilde v_i=\sinh(b_ir-\eta_i)$,
$\eta_i=\zeta_i+\widetilde\zeta_i$ where $\zeta_i$ for $i=3,4$ are
calculated by Eq. (\ref{zetai}) and $\widetilde\zeta_i$ for
$i=1,\ldots,4$ are calculated by the same formula with $a_{1}$ and
$a_{2}$ replaced by $a_{3}$ and $a_{4}$. After some calculations
we obtain an explicit expression for the Wronskian,
\begin{eqnarray}
W(\tilde v_1,\ldots,\tilde v_4)= \nonumber
\sum_{i=1}^3\sum_{j=i+1}^4 (-1)^{i+j} b_ib_j
(b_j^2-b_i^2)\\ \times (b_k^2-b_l^2)
\cosh\xi_i\cosh\xi_j\sinh\xi_k\sinh\xi_l, 
\end{eqnarray}
where $\xi_i=b_ir-\eta_i$ and,
in every term,  $k>l$ take the values from the set
$(1,2,3,4)$ different from the values of $i$ and $j$.
An explicit expression for the obtained potential
is rather involved and we omit it here. Its
typical behavior  is shown on Fig. \ref{fig1}, solid line.

The Jost function for the two resonance potential follows from Eq.
(\ref{F_n}),
\begin{equation}\label{}
F_{2res}=\prod_{j=1}^4\frac{k-\alpha_j}{k+i b_j}\,.
\end{equation}
Thus the resonances occur at $k=k_1=-a_2+i a_1$ and $k=k_2=-a_3+i
a_4$ with the mirror poles at $k=a_2+i a_1$ and $k=a_3+i a_4$.
\begin{figure}[]
\begin{center}
\begin{minipage}{7.5cm}
\epsfig{file=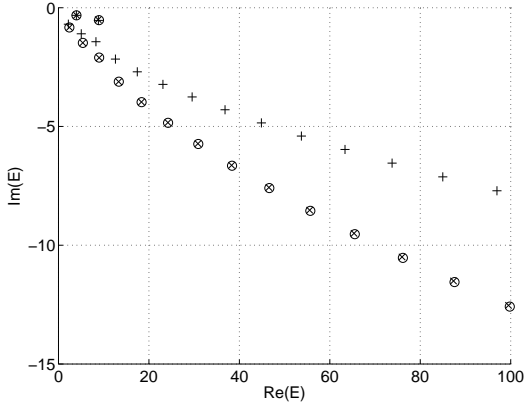, width=7.cm}
\caption{Poles of the truncated two-resonance potential at
$a_1=-0.1$, $a_2=-2$, $a_3=-0.08$, $a_4=-3$, $b_1=0.2$, $b_2=0.1$,
$b_3=0.08$, $b_4=0.05$, and cutoff radius $R_{cut}=5$. The symbols
correspond to the roots of transcendental equation
\eqref{transcen} $(\circ)$, the poles of the $S$ matrix
\eqref{poles} obtained from the eigenvalues of $\Heff$ $(\times)$,
and the fixed-point approximation \eqref{fixed-point_e},
\eqref{fixed-point_g} $(+)$.} \label{fig7}
\end{minipage}
\end{center}
\end{figure}

\begin{figure}[]
\begin{center}
\begin{minipage}{7.5cm}
\epsfig{file=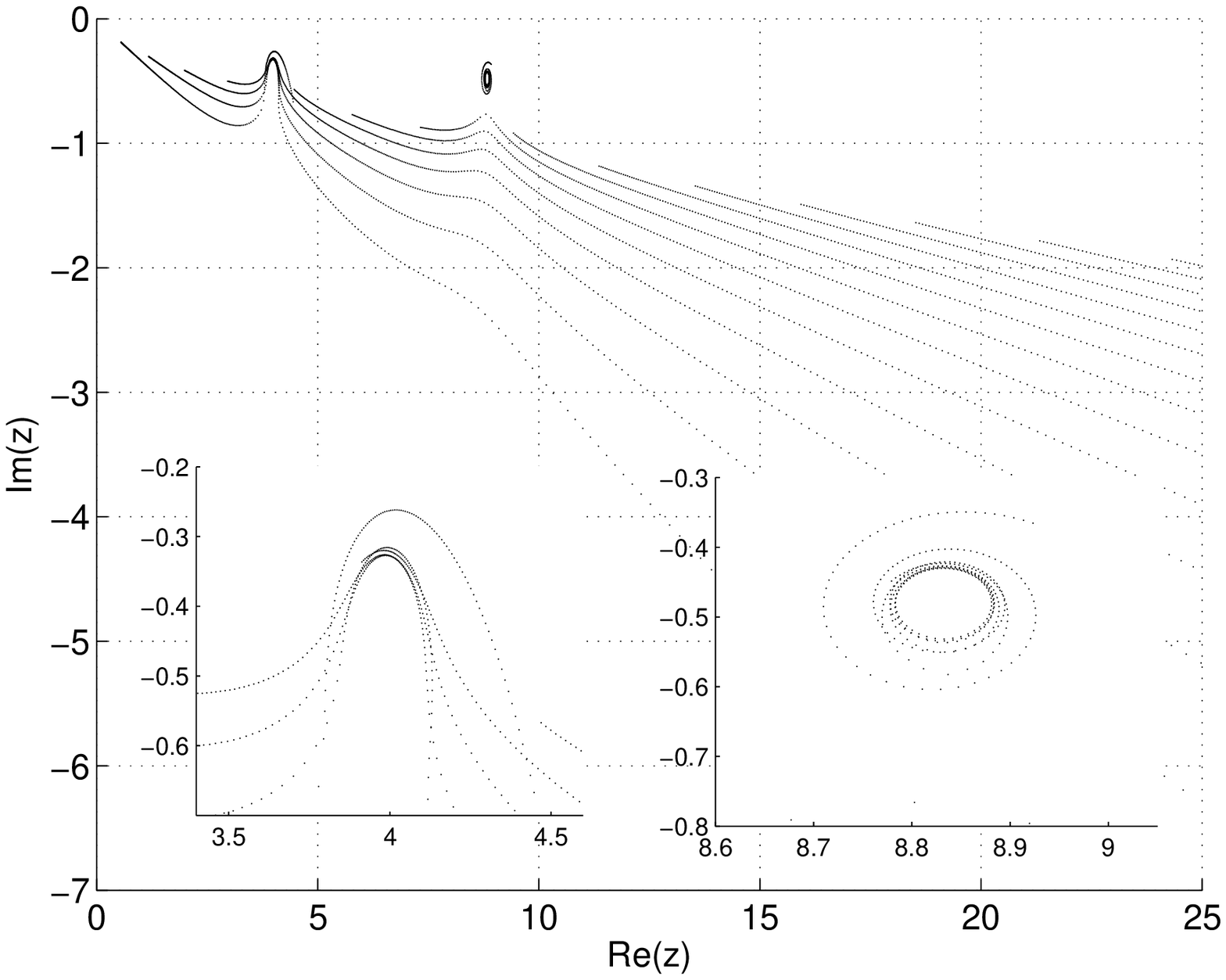, width=7cm}
\caption{Solutions of fixed-point equations \eqref{fixed-point_e}
and \eqref{fixed-point_g} for two-resonance potential $a_1=-0.1$,
$a_2=-2$, $a_3=-0.08$, $a_4=-3$, $b_1=0.2$, $b_2=0.1$, $b_3=0.08$,
$b_4=0.05$. Cutoff radius $R_{cut}$  changes from 0.5 to 7 with
step 0.01. With increasing $R_{cut}$, the trajectories move to
small $\mbox{Re}(z)$, ~$|\mbox{Im}(z)|$ (with the exception of the
spiralling trajectory).} \label{fig8}
\end{minipage}
%
\begin{minipage}{7.5cm}
\epsfig{file=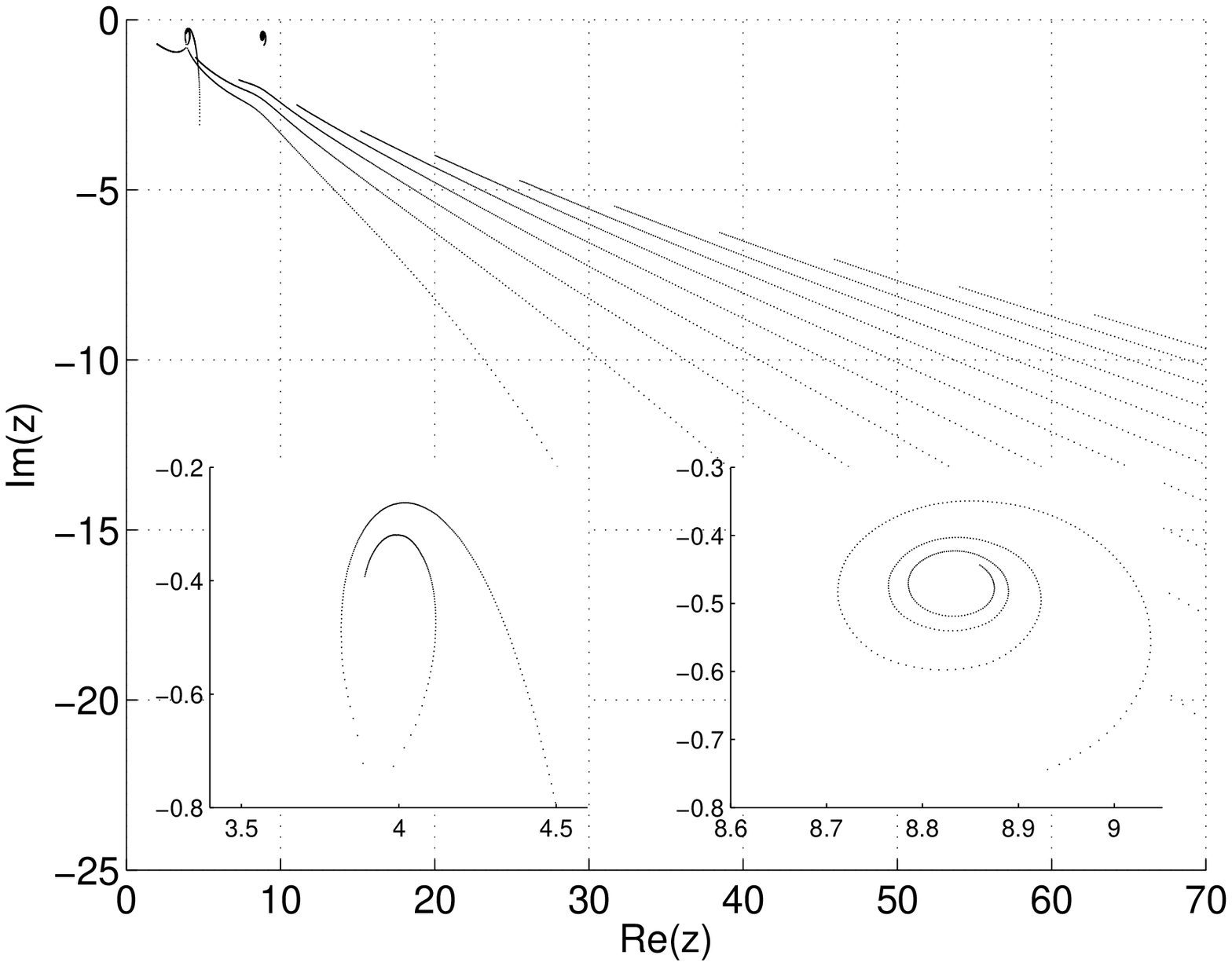, width=7.cm}
\caption{Trajectories of $S$-matrix poles \eqref{poles} for
two-resonance potential at $a_1=-0.1$, $a_2=-2$, $a_3=-0.08$,
$a_4=-3$, $b_1=0.2$, $b_2=0.14$, $b_3=0.08$, and $b_4=0.05$.
Cutoff radius $R_{cut}$ changes from 0.5 to 7 with step 0.01. With
increasing $R_{cut}$,  the trajectories move to small
$\mbox{Re}(z)$,~$|\mbox{Im}(z)|$ (with the exception of the
spiralling trajectory).} \label{fig9}
\end{minipage}
\end{center}
\end{figure}

The resonance behavior of the cross section is more visible when
$S$-matrix poles are close enough to the real axis. This is
achieved by a proper choice of the Bargmann potential parameters.
Results of our calculations are presented in Figs.
\ref{fig7}-\ref{fig9}. For the set of parameters chosen in these
figures the complex energies have the values $E_1=(-a_2+i
a_1)^2=3.99-i 0.4$ and $E_2=(-a_4+i a_3)^2=8.9936-i 0.48$. The
figures show the same features as those obtained for the
one-resonance case. Two narrow resonances stand separately from
the  chain of the cutoff poles. The poles of the $S$ matrix are
determined well enough when calculated according to Eq.
\eqref{poles}. Solutions of the fixed-point approximation
\eqref{fixed-point_e} identify correctly the positions of the
physical poles if they are close enough to the real axis.

\section{Conclusion}

In the present paper, we considered the application of the FPO
formalism to potential scattering in the single-channel case. For
this purpose we analytically constructed model potentials being a
generalization of Bargmann potentials to resonance states with one
and two resonances at given energies.

In the FPO formalism, the corresponding spectroscopic  and
scattering information is obtained from the non-Hermitian
Hamiltonian $\Heff$ and the $S$ matrix derived by means of
$\Heff$. The Hamiltonian $\Heff$ describes the  localized part of
the system under the influence of the coupling to the continuum.
In the present paper, it is obtained in the framework of the
tight-binding model.

First we compared the phase shifts obtained numerically in both
methods. We received an astonishing good agreement of the results
obtained in the two models. The phase shift notifies only the
physical resonances.

To compare the spectroscopic values obtained in the two models, we
are confronted with the problem that the eigenvalues of $\Heff$
involved in the $S$ matrix are energy dependent while the exactly
solvable potentials provide us only the poles of the $S$ matrix.
We therefore have to determine the poles of the $S$ matrix also in
the framework of the FPO formalism. The standard  fixed-point
approximation  \eqref{fixed-point_e} for the positions of the
resonances  is inadequate for this purpose since the widths are
determined by solving Eq. \eqref{fixed-point_g} at the positions
of the resonances. Hence the widths are, generally, erroneous. The
results of our calculations show clearly that the fixed-point
approximation gives, nevertheless, reasonable values if the widths
are small enough.  In order to determine exactly the poles of the
$S$ matrix in the framework of the FPO formalism, we solved the
nonlinear equation \eqref{poles}, ${\rm
Det}\left[E-\Heff\left(E\right)\right]=0$, in the complex $E$
plane.

The results for the spectroscopic values are the following. The
truncation of the potential  in the tight-binding FPO model leads
to the appearance of spurious   solutions of Eq. \eqref{poles}
just as in the well-known case of the $S$-matrix cutoff poles
\cite{Meyer}. The physical resonances of the truncated Bargmann
potentials are well described  by the complex energies satisfying
Eq. \eqref{poles}, i.e., by the $S$ matrix poles calculated in the
FPO formalism. Furthermore, the spurious solutions coincide with
the cutoff poles of the scattering matrix. The last ones behave
differently from the physical resonances in repeated calculations
with different parameter sets. In our case the parameter is the
cutoff radius $R_{cut}$  (the potential is set to zero at
coordinates $r\geq R_{cut}$). The physical poles representing
visual resonances are stable against variation of $R_{cut}$, in a
certain range, in contrast to the cutoff poles that are not
stable.

\acknowledgments{ VVS acknowledges a support from INTAS Grant No
06-1000016-6264. VVS and BFS are partially supported by grants
RFBR-06-02-16719 and SS-871.2008.2. VVS and KNP thank the MPIPKS
Dresden for hospitality.}

\end{document}